\documentclass[nofootinbib,prl,12pt,reprint,longbibliography]{revtex4-2}

\usepackage{graphicx}
\usepackage{epstopdf}
\usepackage{amsmath}
\usepackage{amssymb}
\usepackage{amsfonts}
\usepackage{amsthm}
\usepackage{color}
\usepackage{array}
\usepackage{verbatim}
\usepackage{url}
\usepackage{graphicx}
\usepackage{braket}
\usepackage{epsfig}

\usepackage[
colorlinks=true,
linkcolor=blue,
urlcolor=magenta,
filecolor=blue,
citecolor=red,
pdfstartview=FitV,
pdftitle={},
pdfauthor={},
pdfsubject={},
pdfkeywords={},
pdfpagemode={},
bookmarksopen=true
]{hyperref}
\usepackage{lipsum}
\usepackage{lipsum}

\usepackage{caption, subcaption}

\def\tanh{{\mathrm{tanh}}}

\def\frac#1#2{{#1\over #2}}

\def\p{\partial}

\def\d{{\mathrm{d}}}

\def\be{\begin{equation}}
	\def\ee{\end{equation}}
\def\ba{\begin{eqnarray}}
	\def\ea{\end{eqnarray}}

\usepackage{titlesec}
\titleformat{\paragraph}[runin]
{\normalfont\scshape\bfseries}{\theparagraph}{1em}{}

\begin{document}
	
	\title{Can we detect a supertranslated black hole?}

	\author{Subhodeep Sarkar}\email{subhodeep.sarkar1@gmail.com }
	\affiliation{Indian Institute of Information Technology (IIIT), Allahabad, \\
		Deoghat, Jhalwa, Uttar Pradesh, India 211015}
	
	\author{Shailesh Kumar}\email{shaileshkumar.1770@gmail.com }
	\affiliation{Indian Institute of Information Technology (IIIT), Allahabad, \\
		Deoghat, Jhalwa, Uttar Pradesh, India 211015}
	
	\author{Srijit Bhattacharjee}\email{srijuster@gmail.com}
	\affiliation{Indian Institute of Information Technology (IIIT), Allahabad, \\
		Deoghat, Jhalwa, Uttar Pradesh, India 211015}
	
	\date{\today}
	
	\begin{abstract}
		We attempt to address the question of whether standard tests of general relativity can detect the presence of a black hole carrying a supertranslation field. In this regard, we study the photon sphere of a dynamical black hole carrying a supertranslation hair. We find that the dynamics of the photon sphere is quite subtle and it may offer an opportunity to differentiate a supertranslated black hole from its bald counterpart. This represents a first step towards understanding the observational signatures of a supertranslated dynamical black hole. 
	\end{abstract}

	\maketitle
	
	\paragraph{Introduction.}
	Since the direct detection of gravitational waves \cite{PhysRevLett.116.061102, PhysRevLett.125.101102}, testing different properties and hypotheses related to black hole (BH) spacetimes has become a rapidly growing field of research. The image of the \emph{shadow} of a black hole produced by the Event Horizon Telescope (EHT) Collaboration \cite{EventHorizonTelescope:2019dse, EventHorizonTelescope:2019uob, EventHorizonTelescope:2019jan, EventHorizonTelescope:2019ths, EventHorizonTelescope:2019pgp, Doeleman2008} is another recent groundbreaking result in black hole physics.  The BH shadow is caused by the strong gravitational lensing of light near the \emph{photon sphere}.  It is essentially a blind spot in the field of view of an observer; we cannot receive any photon from that region since all photon trajectories lying inside the photon sphere inevitably spiral into the black hole. There has been a lot of effort in understanding the shadow cast by various kinds of stationary black holes in recent years (see \cite{Cunha:2018acu,Perlick:2021aok} and the references therein). Notably, the shadow due to dynamical black holes was explored in \cite{Mishra:2019trb}.
	
	Black holes possess very few parameters that can be used to characterize them, for example, mass ($M$), charge ($Q$) and angular momentum ($J$). According to the Uniqueness theorem \cite{Israel1967, Robinson1975}, for asymptotically flat (AF) black holes, $M, Q, J$ are the only three conserved charges that we can get along with the other Poincar\'{e} charges. It is also known from the seminal work by Bondi, Metzner, van der Burg and Sachs (BMS) \cite{doi:10.1098/rspa.1962.0161, Sachs}, that the symmetry group of AF spacetimes gets enhanced at the null infinities. This enhanced symmetry group, known as the BMS group, forms an algebra which is expressed as a semi-direct sum of the Lorentz algebra and an abelian normal subalgebra containing \emph{supertranslations}. These new kinds of symmetries are angle-dependent translations, and they give rise to an infinite number of conserved charges \cite{PhysRevLett.116.231301, Hawking2017, Strominger2016, strominger2018lectures}. In the presence of black holes, the conservation of such charges would mean that the future event horizon of a black hole will also have a contribution to the conservation. Therefore the event horizon should be endowed with supertranslation hairs. This picture has paved a way to address the black hole information paradox. It is believed that the infinite number of (soft) hairs which correspond to the supertranslation charges may provide a mechanism for retrieving the information from an evaporating black hole \cite{PhysRevLett.116.231301, Strominger2016}. The BMS group has recently been further extended to accommodate another set of symmetries called \emph{superrotation} \cite{PhysRevLett.105.111103, barnich2012supertranslations, Compere:2018aar}. In principle, different supertranslated black holes are distinguishable via their (supertranslation field dependent) superrotation charges.
	
	The existence of supertranslation (and superrotation) hairs can be verified if one considers gravitational waves carrying such fields as they pass through stationary detectors situated near the asymptotic infinities. The passing of such waves induces permanent changes to the detectors and the clocks leading to gravitational memory effects \cite{Strominger2016, Pasterski2016, PhysRevD.101.083026, Grant:2021hga, PhysRevLett.117.061102}. It is believed that the memory effect will be detectable in advanced gravitational wave detectors, although the connection between such a possible detection and BMS symmetries is not very clear \cite{Favata:2010zu, Favata:2009ii, Favata:2008yd, Favata:2011qi, PhysRevD.103.044012}.
	
	The possibility of the existence of BMS-like symmetries near the event horizon of a black hole has generated a lot of activity recently, and there are several ways by which BMS symmetries have been recovered near the black hole horizon \cite{PhysRevLett.116.091101, Donnay2016, Compere2016, Compere:2016hzt,Ashtekar2018, Blau2016, PhysRevD.98.104009, Chandrasekaran2018, Iofa:2017ukq,Iofa:2018pnf, Iofa:2019jjm}. Finding signatures of such symmetries in a way analogous to those which are found in the far region should be an obvious way to verify the whole concept. The memory effect on a near horizon asymptotic metric carrying supertranslation-like fields has been studied in \cite{PhysRevD.98.124016}. An explicit relation between the displacement memory effect and BMS-like symmetries has been considered in \cite{Bhattacharjee2021, PhysRevD.102.044041, PhysRevD.100.084010}. 
	
	A memory tensor for black hole horizons has also been proposed in \cite{Rahman:2019bmk}, and it may be used as a useful device to detect supertranslation memory.  This memory tensor is expressed in terms of the expansion and the shear corresponding to the null congruence that generates a null horizon or a null boundary \cite{Rahman:2019bmk, Grumiller:2020vvv, Adami:2021nnf} of a spacetime. This way of presenting the memory takes into account a dynamic era of a black hole horizon sandwiched between two stationary eras. The dynamic era will induce a permanent relative displacement of the null geodesics of the horizon and this change will get reflected in the memory tensor. If the null horizon is slightly perturbed from stationarity, then the memory tensor reads  \cite{Rahman:2019bmk}:
	
	\begin{align}\label{memoryt}
		\Delta_{AB}={1\over 2}\int_{v_i}^{v_f}\vartheta^{(k)}\gamma_{AB}\d v+2\int_{v_i}^{v_f}\sigma_{AB}^{(k)}\d v,
	\end{align}
	where $k$ is the horizon generating null vector\footnote{We have used indices $a, b,\cdots$ for 4D spacetimes, $i,j,k\cdots$ for any 3-hypersurface, and $A,B, \cdots$ for a codimension 2 surface.}. Here the early time stationary era is represented by $v<v_{i}$, whereas the late time stationary era is represented by $v>v_{f}$ ($v$ is a null coordinate adapted to the horizon). $\sigma^{(k)}_{AB}$ and $\vartheta^{(k)}$, respectively, denote the shear and expansion with respect to the rotation-free null congruence generated by $k$. Further, $\gamma_{AB}$ is the codimension-1 metric to the null surface whose Lie derivative along $k$ is related to the second fundamental form $\mathcal{K}_{AB}$, and yields the following form
	\begin{align}\label{lieDgm}
		\mathcal{L}_{k}\gamma_{AB} = 2\mathcal{K}^{(k)}_{AB} = \frac{2}{(D-2)} \vartheta^{(k)}\gamma_{AB}+2\sigma^{(k)}_{AB}.
	\end{align}
	It is apparent that this memory is related to the change in $\gamma_{AB}$ between two stationary eras. In this context, it is necessary to generalize the memory tensor to dynamical situations as well, given that those are practically more relevant. To do so, we consider a metric containing a dynamical horizon, the supertranslated Vaidya black hole (STVBH). Such a configuration can result from a Vaidya solution subjected to a linear perturbation resulting from a shock-wave like energy-momentum flux \cite{Hawking2017,strominger2018lectures,Chu2018}.
	We can solve the Einstein field equation with the following stress-energy tensor, 
	
	\begin{align}
		T_{\mu \nu}^{\mathrm{tot}} =    \bar{T}_{\mu \nu} +  T_{\mu \nu}, \label{stressenergy}
	\end{align}
	where the non-zero components of the above tensors are
	\begin{subequations}
		
		\begin{align}
			\bar{T}_{v v} = & \dfrac{\dot{M}(v)}{4 \pi r^2}, \\
			T_{vv} = &  \dfrac{1}{4 \pi r^2}\left( \hat{\mu}(v, \theta) + \hat{T}(\theta) \delta(v-v_0) \right) \nonumber \\
			& + \dfrac{1}{4 \pi r^3}\left( \hat{T}^{(1)}(\theta)\delta(v-v_0) + \hat{t}^{(1)} \Theta(v-v_0)\right), \\
			T_{v \theta} = & \dfrac{1}{4 \pi r^2}\left( \hat{T}_{\theta}(\theta)\delta(v-v_0) + \hat{t}_{\theta} \Theta(v-v_0)\right),
		\end{align}
	\end{subequations}
	with
	\begin{align*}
		& \hat{\mu}(v, \theta) = \partial_v \left( \Theta(v-v_0)(f(\theta) \dot{M}(v) + \mu)\right), \\
		& \hat{T}(\theta) = -\dfrac{1}{4} D^2(D^2+2)f(\theta),  ~ \hat{T}^{(1)} = \dfrac{3M(v)}{2} D^2f(\theta), \\
		& \hat{T}_{\theta} = \dfrac{3M(v)}{2} D_\theta f(\theta), ~  \hat{t}_{\theta} = \dot{M}(v) D_\theta f(\theta), \\
		& \hat{t}^{(1)} = \dot{M}(v) D^2f(\theta). \label{stvbh_tmunu}
	\end{align*}
	Here, a dot denotes the derivative with respect to $v$, $\delta(v-v_0)$ and $\Theta(v-v_0)$ are the delta function and step function respectively, $\mu$ is a constant, $f(\theta)$ is an arbitrary function of $\theta$ that represents the supertranslation field, and $D$ is the covariant derivative with respect to the metric on the unit two sphere. The $T_{\mu \nu}$ prescribed above represents a shockwave of strength $\mu$ that implants the supertranslation hair to a Schwarzschild Vaidya black hole \cite{Chu2018}, and the resultant metric reads,
	\begin{subequations}
		\begin{align}
			\d s^2&=\d \bar{s}^2 + h_{\mu \nu} \d x^{\mu} \d x^{\nu} \\ &= -g_{vv} \d v^2 + 2 \d v \d r \nonumber \\ & + g_{v \theta} \d v \d \theta + r^2 \tilde{g}_{\theta \theta} \d \theta^2 + r^2 \sin^2 \theta \tilde{g}_{\phi \phi} \d \phi^2, \label{stvbh}
		\end{align}\label{STVBH}
	\end{subequations}
	
	where $\d \bar{s}^2=\bar{g}_{\mu\nu}dx^{\mu}dx^{\nu}$ is the Schwarzschild-Vaidya metric with mass parameter $M(v)$, $h_{\mu \nu}$ is a perturbation given by
	\begin{align}
		h_{\mu \nu} = \Theta(v-v_0)\left(\mathcal{L}_{\chi_{f}} \bar{g}_{\mu\nu} + \dfrac{2 \mu}{r} \delta^{v}_{\mu} \delta^{v}_{\nu}\right),
	\end{align}
	where the Lie derivative is carried out with respect to the Killing vector field
	\begin{align}
	    \chi_{f} = f(\theta)\partial_{v}-\frac{1}{2}D^{2}f(\theta)\partial_{r}+\frac{1}{r}D^{A}f(\theta)\partial_{A}. 
	\end{align}
	The metric components are,
	\begin{subequations}
		\begin{align} 
			g_{vv}  = & ~ 1-\frac{2 M(v)}{r}-\Theta(v-v_0)\left(\frac{M(v)}{r^2}f''  \right.\nonumber \\
			& \left. +\dfrac{M(v)\cot \theta}{r^2}f'+\frac{2 \dot{M}(v)}{r} f \right),\\
			g_{v \theta}  = & ~ \Theta(v-v_0)\left( \csc^2 \theta f' - 2\left(1-\dfrac{2 M(v)}{r}\right)f' \nonumber \right. \\ & \left. - \cot \theta f'' -f''' \right.\Big{)},\\
			\tilde{g}_{\theta \theta} = & ~ 1 + \Theta(v-v_0)  \left(\dfrac{f''}{r} -\dfrac{\cot \theta}{r } f' \right), \\
			\tilde{g}_{\phi \phi} = & ~ 1 - \Theta(v-v_0) \left(\dfrac{f''}{r} - \dfrac{\cot \theta}{r } f' \right).
		\end{align}
	\end{subequations}
	Here, prime denotes derivative with respect to $\theta$. The supertranslation $f$ is assumed to be small and it depends only on the $\theta$ coordinate. Since the perturbations are not spherically-symmetric, the metric is  not spherically symmetric as well. We also assume $M+\mu \simeq M$. This metric has a dynamical horizon ($H$), located at 
	\begin{align}
		r_H=2 M(v) +2 f \dot{M}(v)+ \dfrac{1}{2}(f''+f' \cot \theta) +\mathcal{O}(f^2).    
	\end{align}
	In the double null foliation of $H$, there will be two null normals $(k,l)$ to the codimension one hypersurface $H$. In this case, the expansion $\tilde{\vartheta}^{(k)}$ will vanish as $k $ is the null-normal to the horizon $H$. The other expansion $\tilde{\vartheta}^{(l)}$ will not be zero and it will have a negative value \cite{Chu2018}. Therefore $H$ is a supertranslated version of a Future Outer Trapping Horizon (FOTH) \cite{Ashtekar:2003hk, Ashtekar2004, Chatterjee:2020enf, Ghosh:2020wjx}. Since $\tilde{\vartheta}^{(k)}$ is zero, and $\tilde{\vartheta}^{(l)}$ is independent of supertranslation hair on $r_H$, we do not get any memory. In this regard, there could be an equivalent way to define the memory tensor. We can consider the congruence generated by the spacelike normal to the codimension one hypersurface of $H$ given by $r^a=\frac{1}{\sqrt{2}}(k^a-l^a)$ \footnote{$k^a=\{1,\frac{g_{vv}}{2},0,0\};~~  l^a=\{0,-1,0,0\}$} . As $H$ is not null, we have a slight change in the expression of memory tensor written above in (\ref{memoryt}). 
	\begin{align}\label{memoryt}
		\Delta^{(DH)}_{ij}={1\over 3}\int_{v_i}^{v_f}\vartheta^{(r)}\gamma_{ij}\d v+\int_{v_i}^{v_f}\sigma_{ij}^{(r)}\d v +\int_{v_i}^{v_f}\omega^{(r)}_{ij}\d v.
	\end{align}
	An analogous expression can be written as \eqref{lieDgm} for the  Lie derivative of the induced metric on the dynamical horizon, $\gamma_{ij}$, with respect to the spacelike normal. The quantities $\sigma^{(r)}_{ij}$, $\vartheta^{(r)}$ and $\omega^{(r)}_{ij}$ represent the shear, expansion and rotation respectively for $r^{a}$.
	The expansion for the congruence generated by $r^a$ is shown below and it includes the supertranslation field. 
	\begin{align}
		\vartheta^{(r_{H})} =& \frac{1}{4\sqrt{2}M(v)^{2}}\Big(5(M(v)-f\dot{M}(v))-D^{2}f\Big)+ \mathcal{O}(f^2).
	\end{align}
	
	Note that the memory tensor $\Delta^{(DH)}_{ij}$ contains a rotation part also. The only non-vanishing component of the rotation will be $\omega_{v\theta}.$ There will be a non-zero shear part as well. Interestingly, only the rotation vanishes if we set $f$ to be a constant on $H$ (a non-supertranslated horizon) \cite{Chatterjee:2020enf, Ghosh:2020wjx}.  The motion along the vector $r^a$ can be regarded as the time evolution of $H$ for a distant observer.
	
	
	Now, a pertinent question is whether such memories can be detected through experiments performed near Earth. Investigating these memories requires one to place the detectors near the horizon of a black hole. This seems to be a far fetched idea. A possible approach could be to see if a supertranslated configuration can be detected through the standard tests of GR like gravitational lensing or by studying the black hole shadow. In this context, let us first study the photon sphere of a supertranslated Schwarzschild black hole \cite{Hawking2017} whose line element $\d {s}_0^2$ can be recovered by setting $M(v)=$ constant in \eqref{STVBH}.

	The photon sphere of a black hole is an unstable null geodesic that forms a close orbit around it. It acts as a boundary that separates all the photons reaching the black hole into either fly-by orbits (which escape to infinity) or capture orbits (which spiral into the black hole) depending on the value of the impact parameter with which the photons approach the black hole \cite{Perlick:2021aok}. This gives rise to the prospect of detecting a black hole by observing its so-called shadow. In our study, we shall consider only those photon orbits which lie in the equatorial plane and therefore set $\theta= \pi/2$. This essentially amounts to projecting the photon sphere onto the equatorial plane and studying the photon circle.  For a static (or stationary) black hole, the radius of the photon circular orbit $r_0$ is constant.  For example, it is $3 M$ for a Schwarzschild black hole whereas a Kerr black hole has two photon circular orbits of constant radii in the equatorial plane which depend on both the mass and spin parameter of the BH \cite{1972ApJ...178..347B}. We now show that the photon sphere of a supertranslated Schwarzschild black hole differs from its bald counterpart. 
	
	For a null trajectory in the supertranslated Schwarzschild spacetime, $\d {s}_0^2=0$. If we then consider a photon circular orbit in the equatorial plane, in this static spacetime, we can further set $r=r_0$, a constant, and obtain,
	\begin{align}
		& \left( \dfrac{\d \phi}{\d v}\right)^2 =  \dfrac{1}{r_0^2} \label{nullity1} \dfrac{g_{vv}}{\tilde{g}_{\phi \phi}} \Bigr{|}_{r=r_0, \theta=\pi/2}. 
	\end{align}
	Now, we can also calculate the radial geodesic equation in the equatorial plane for a null trajectory  and write
	\begin{align}
		&   2 \dfrac{ \d^2 r}{\d \lambda^2} + g_{vv}\dfrac{\partial g_{vv}}{\partial r} \left(\dfrac{\d v}{\d \lambda}\right)^2  - 2 \dfrac{\partial g_{vv}}{\partial r} \left(\dfrac{\d v}{\d \lambda}\right)\left(\dfrac{\d r}{\d \lambda}\right) \nonumber \\
		& - g_{vv}\dfrac{\partial(r^2 \tilde{g}_{\phi \phi})}{\partial r}\left(\dfrac{\d \phi}{\d \lambda}\right)=0.
	\end{align}
	where $\lambda$ is an affine parameter associated with the null geodesics.	If we now set $r=r_0$, and consequently $\dot{r}=0$ and $\ddot{r}=0$ in the above radial geodesic equation, we will obtain,
	
	\begin{align}
		\left(\dfrac{\d \phi}{\d v}\right)^2 = \dfrac{\partial_r g_{vv}}{\partial_r (r^2 \tilde{g}_{\phi \phi})}\Bigr{|}_{r=r_0, \theta=\pi/2}. \label{radial1}
	\end{align}
	We can then equate the right hand sides of \eqref{nullity1} and \eqref{radial1} to get an algebraic equation that can be solved to obtain the radius of the photon sphere of a supertranslated Schwarzschild black hole, viz,
	\begin{align}
		r_0 = 3 M + \dfrac{1}{2} f''(\theta)|_{\theta = \pi/2} + \mathcal{O}(f^2) \label{bc}
	\end{align}
	
	However, from the perspective of detection, a static or stationary bald black hole may not be distinguishable from its supertranslated counterpart, as an observer performing a local experiment may not have any clue whether they are just two different black holes with two different masses (or angular momenta) or if they differ by a supertranslation hair. On the other hand, practically one does not find an ideal static black hole in nature. Black holes interact with the surroundings and become dynamic. Therefore we need to consider a dynamical black hole that might not be plagued with such limitations.
	
	\paragraph{Photon Sphere of a Supertranslated Vaidya black hole.}We now study the evolution of the photon sphere of an STVBH given by \eqref{STVBH} following the method elucidated in \cite{Mishra:2019trb}. 
	
	We add a perturbation of the form $\frac{2\epsilon h(v)}{r} d \theta d\phi$ to the metric \eqref{stvbh} with $\epsilon \ll 1$. Since we are interested in photon circular orbits in the equatorial plane, our analysis is not sensitive to this small deformation. The perturbed STVBH is obtained by adding the following stress-energy tensor, $T^{\mathrm{pert}}_{\mu \nu}$,  to \eqref{stressenergy} where
	\begin{align}
		T^{\mathrm{pert}}_{(\mu \nu)}   = \dfrac{\tau_1}{8 \pi r^3} \delta^{\phi}_{(\mu} \delta^{v}_{\nu)} + \dfrac{\tau_2}{8 \pi r^4} \delta^{\phi}_{(\mu} \delta^{r}_{\nu)} + \dfrac{\tau_3}{8 \pi r^4} \delta^{\phi}_{(\mu} \delta^{\theta}_{\nu)},
	\end{align}
	with $(\cdots)$ denoting symmetrization of the índices $\mu,\nu$ and
	\begin{align}
		&   \tau_1 = \epsilon h'(v) \cot \theta, \nonumber \\
		& \tau_2 = -3 \epsilon h(v) \cot \theta, \nonumber \\
		&   \tau_3 = 2\epsilon(9h(v)M(v)-3rh(v)+2r^2h'(v)).
	\end{align}
	The function $h(v)$ is rather arbitrary but it can be modeled as Gaussian function so that it is absent at very early and very late times. During the dynamic phase, this perturbed black hole is not diffeomorphic to the Vaidya solution. It can be shown that $T^{\mathrm{pert}}_{\mu \nu} k^{\mu} k^{\nu}=0$ and that $T_{\mu \nu}^{\mathrm{tot}}$ satisfies the null energy condition.
	
	The radius of the photon sphere, $r_0$, of a dynamical black hole is not constant. Instead, it becomes a function of time, $r_0 = r_0 (v)$. The evolution of the photon sphere is then governed by a second-order differential equation which we now derive. 
	
	Since $r_0=r_0(v)$, we can write,
	\begin{align}
		&	{\d r_0(v)}  =  \dfrac{\p r_0(v)}{\p v} \d v = \dot{r}_0 \d v.\label{rv}
	\end{align}
	As before, we are considering null orbits in the equatorial plane. So we have $\d s^2 =0$, which along with \eqref{rv} gives us
	\begin{align}
		\left(\dfrac{\d \phi}{\d v}\right)^2 =   h_1(r_0(v), v) + \dot{r}_0 h_2(r_0(v),v), \label{Heq}
	\end{align}
	where 
	$h_1 = {g_{vv}}/({r_0^2(v) \tilde{g}_{\phi \phi}})$, and $h_2 = - {2}/({r_0^2(v) \tilde{g}_{\phi \phi}})$.
	Moreover, in the equatorial plane, the relevant geodesic equations tell us
	\begin{subequations}
		\begin{align}
			\dfrac{\d^2 v}{ \d \lambda^2} &= F_1 (r, v)\left( \dfrac{\d v}{\d \lambda}\right)^2 + F_2 (r, v)\left( \dfrac{\d \phi}{\d \lambda}\right)^2 \label{Feq} ,\\ 
			\dfrac{\d^2 r}{ \d \lambda^2} &=  G_1 (r, v)\left( \dfrac{\d r}{\d \lambda}\right)^2 + G_2 (r, v)\left( \dfrac{\d \phi}{\d \lambda}\right)^2  \nonumber \\
			& + G_3 (r, v) \left(\dfrac{\d v}{\d \lambda}\right)\left( \dfrac{\d r}{\d \lambda}\right) + G_4 (r, v)\left( \dfrac{\d v}{\d \lambda}\right)^2, \label{Geq}
		\end{align}
	\end{subequations}
	where the function $F_i$'s and $G_i$'s depend on the metric components and their derivatives and they can be read off directly from the geodesic equations.
	
	Further, \eqref{rv} implies
	\begin{align}
		\dfrac{\d^2 r_0(v)}{ \d \lambda^2}  = \dot{r}_0(v) \dfrac{\d^2 v}{ \d \lambda^2} + \ddot{r}_0(v)\left( \dfrac{\d v}{\d \lambda}\right)^2. \label{psp1}
	\end{align}
	Now by setting $r=r_0(v)$ and substituting \eqref{Feq}, \eqref{Geq} in \eqref{psp1} together with \eqref{Heq}, we arrive at the second-order differential equation governing the evolution of the photon sphere, viz.,
	\begin{align}
		&	\ddot{r}_0(v) + \dot{r}_0(v)\left[ f_1 + f_2 h_1 -g_2 h_2 -g_3 \right] \nonumber \\
		& +  \dot{r}^2(v)[f_2 h_2 - g_1] 	-\left[g_2 h_1 + g_4\right] =0, \label{psp2}
	\end{align} 
	where $f_i(r_0(v), v) =  F_i(r,v)|_{r=r_0(v)}$ and $g_i(r_0(v), v) =  G_i(r,v)|_{r=r_0(v)}$. 
	
	For concreteness, we can assume that the supertranslation field $f(\theta)$ is of the form,
	\begin{align}
		f(\theta) = q P_2 (\cos(\theta)) \label{P2},
	\end{align}
	where $P_2(\cos \theta)$ is the second Legendre polynomial and $q$ is a constant such that $0<q \leq 1$ (and $f(\theta)$ has been normalized with respect to $M_0$). Note that the supertranslation field and its derivatives become constant on setting $\theta = \pi/2$, so $q$ essentially characterizes the \emph{strength} of the supertranslation field and hence can be called the supertranslation hair. We also restrict ourselves to small values of $q$. So, up to the first order in $q$, the differential equation \eqref{psp2} for the STVBH reads,
	\begin{widetext}
		\begin{align}
			&  \ddot{r}(v) +  \frac{\dot{M}(v)}{r(v)}-\frac{9 M(v) \dot{r}(v)}{r(v)^2}-\frac{6 M(v)^2}{r(v)^3}+\frac{5 M(v)}{r(v)^2} +\frac{3 \dot{r}(v)}{r(v)}-\frac{2 \dot{r}(v)^2}{r(v)}-\frac{1}{r(v)} + q \Theta(v-v_0) \left(\frac{9 \dot{M}(v) \dot{r}(v)}{2 r(v)^2}  -\frac{\ddot{M}(v)}{2 r(v)} \right. \nonumber \\
			& \left. +\frac{6 M(v) \dot{M}(v)}{r(v)^3}-\frac{\dot{M}(v)}{r(v)^2}  -\frac{27 M(v) \dot{r}(v)}{r(v)^3}  -\frac{27 M(v)^2}{r(v)^4}+\frac{15 M(v)}{r(v)^3}+\frac{9 \dot{r}(v)}{2 r(v)^2}-\frac{3 \dot{r}(v)^2}{r(v)^2}-\frac{3}{2 r(v)^2} \right) \nonumber \\
			& + q \delta(v-v_0)\left(\frac{3}{2 r(v)} -\frac{\dot{M}(v)}{2 r(v)}-\frac{3 M(v)}{2 r(v)^2}-\frac{3 \dot{r}(v)}{r(v)} \right) + \mathcal{O}(q^2)= 0.  \label{goryequation}
		\end{align}
	\end{widetext}

	
	\begin{figure}
		\centering
		\begin{subfigure}[h]{0.45\textwidth}
			\centering
			\includegraphics[scale=0.40]{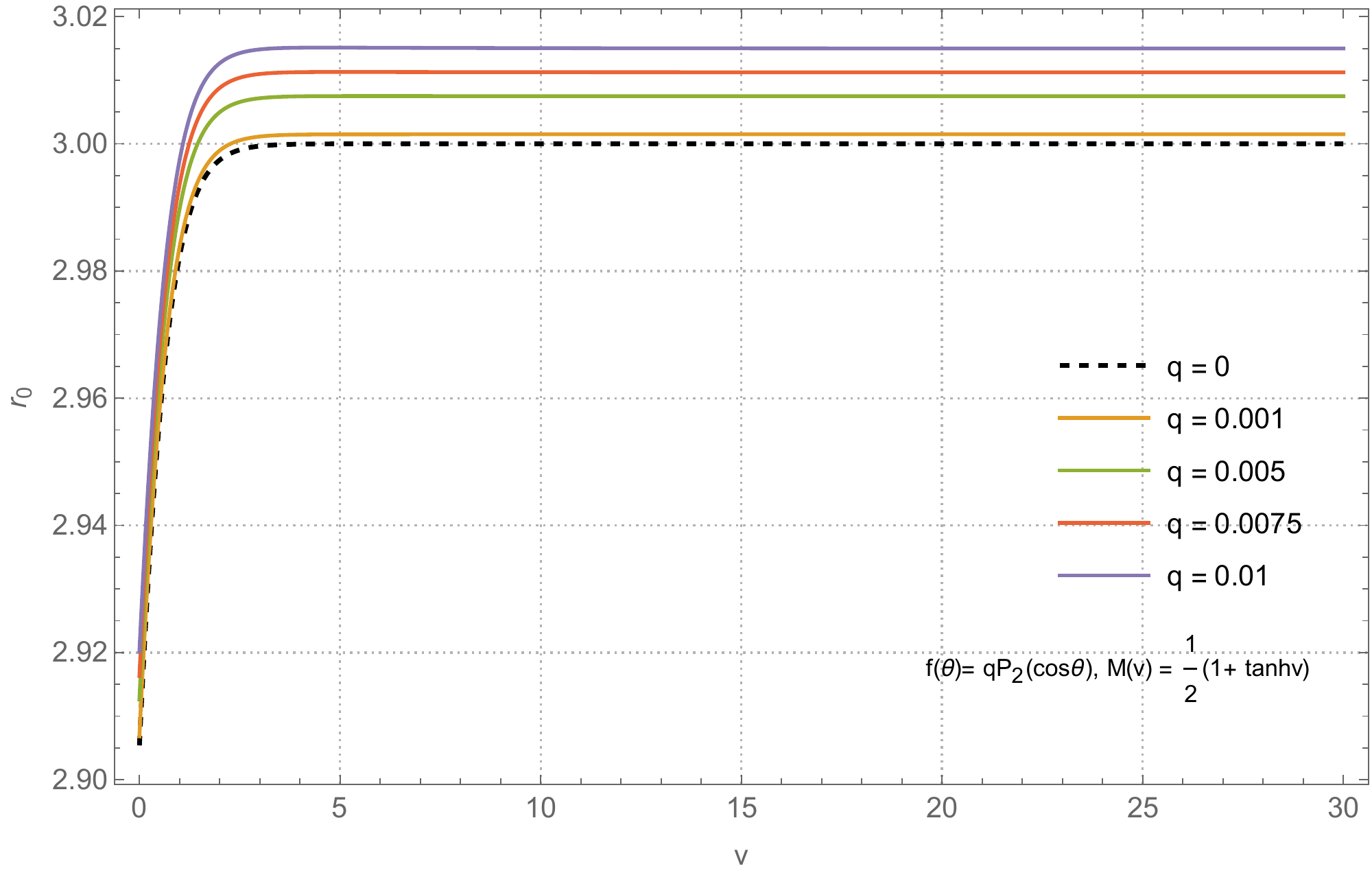}
			\caption{The evolution of the photon sphere for different values of supertranslation hair $q$ from an instant $v>v_0$  with $ M_0=1$.}
			\label{fig:1a}
		\end{subfigure}%
		\vspace{2mm}
		\begin{subfigure}[h]{0.45\textwidth}
			\centering
			\includegraphics[scale=0.40]{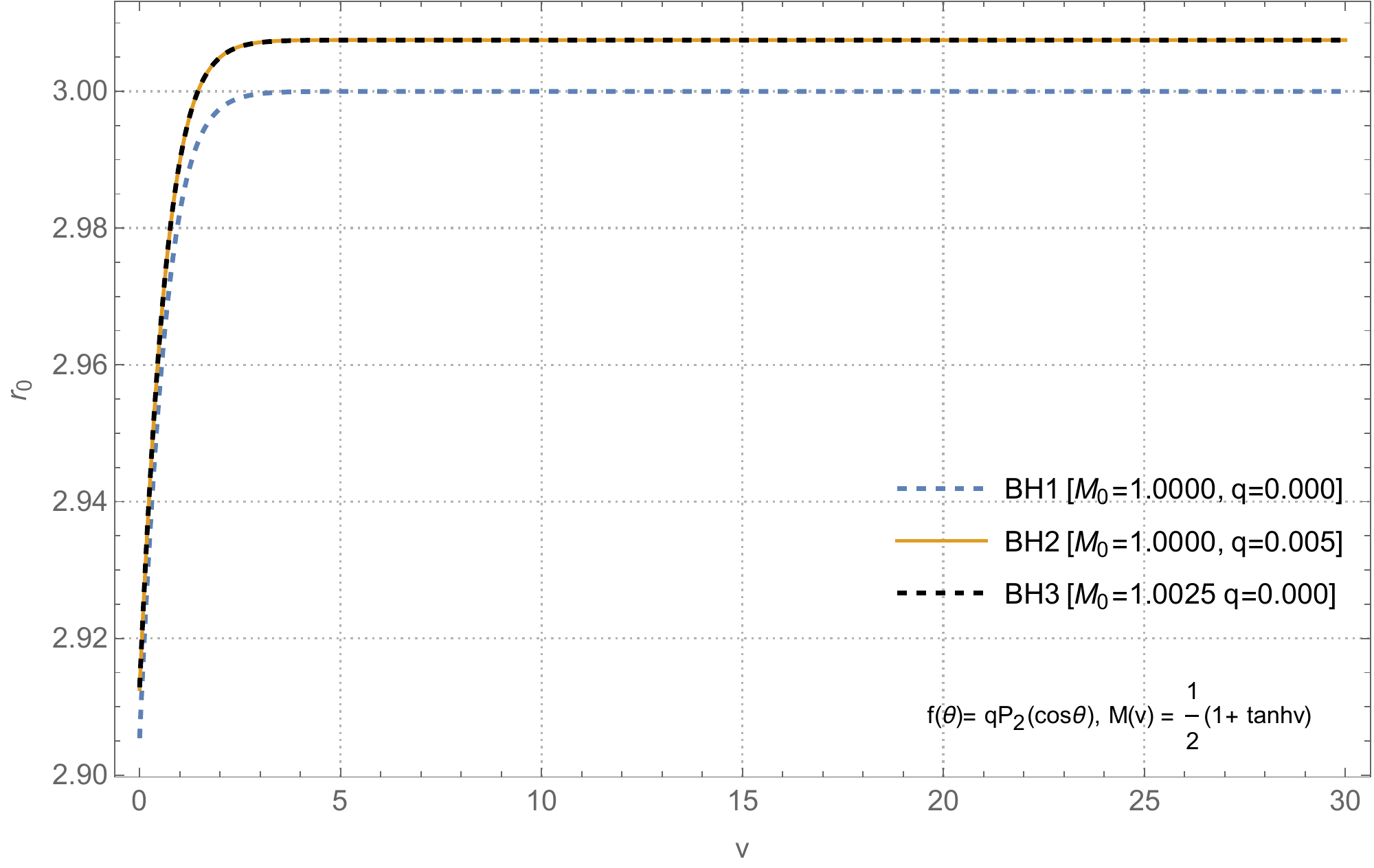}
			\caption{The evolution of three distinct BH configurations from an instant $v>v_0$.}
			\label{fig:1b}
		\end{subfigure}
		\caption{The evolution of the photon sphere of a STVBH from an instant of time $v>v_0$ (i.e., after the implantation of the supertranslation hair by a null shockwave) with $M(v)=\dfrac{1}{2}(1+ \tanh(v)), v_f=30$.}
		\label{STVBH1FIG}
	\end{figure} 
	\begin{figure}
		\centering
		\begin{subfigure}[h]{0.45\textwidth}
			\centering
			\includegraphics[scale=0.40]{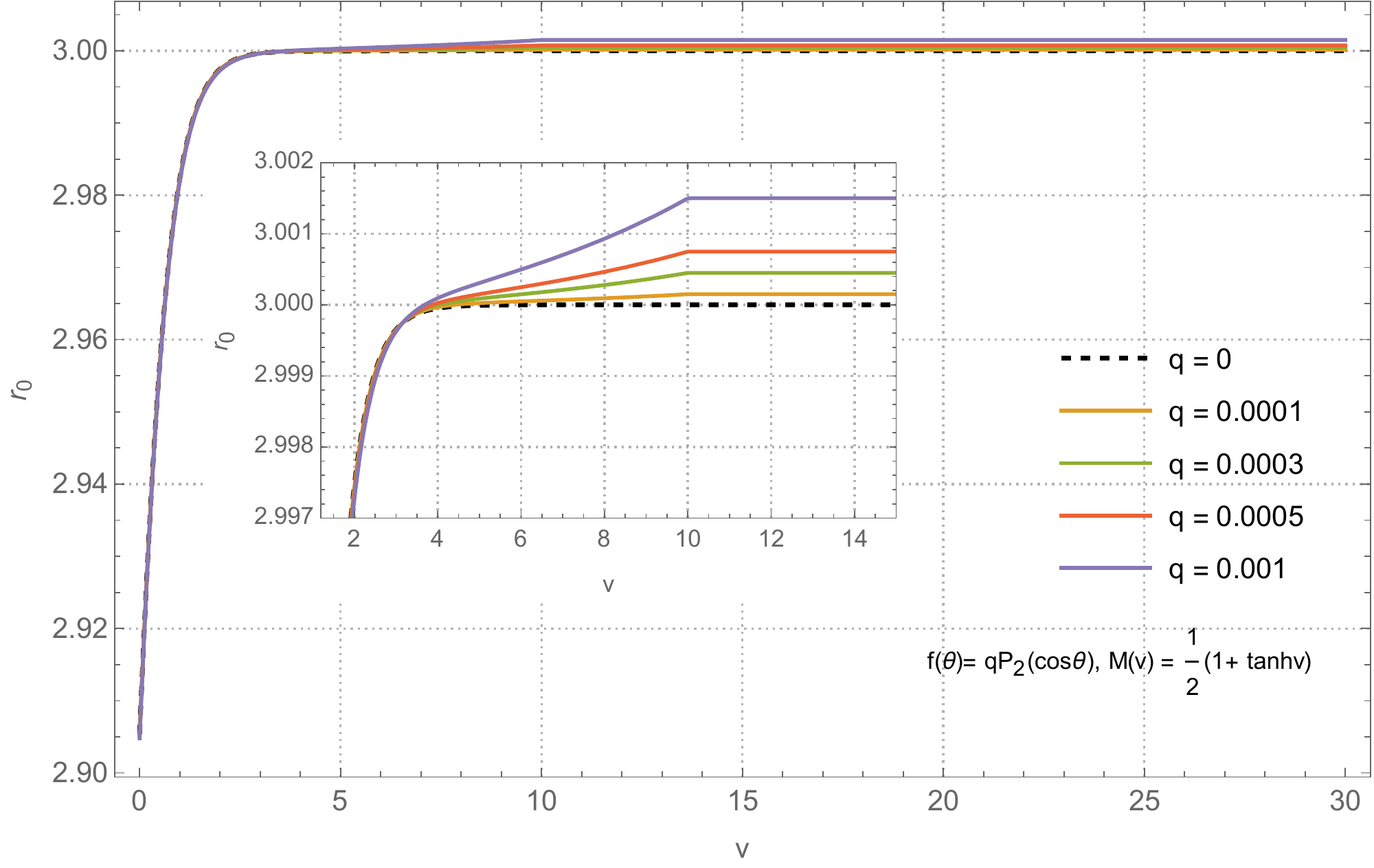}
			\caption{The evolution of the photon sphere for different values of supertranslation hair $q$ from an instant $v<v_0$  with $ M_0=1$.}
			\label{fig:2a}
		\end{subfigure}%
		\vspace{2mm}
		\begin{subfigure}[h]{0.45\textwidth}
			\centering
			\includegraphics[scale=0.40]{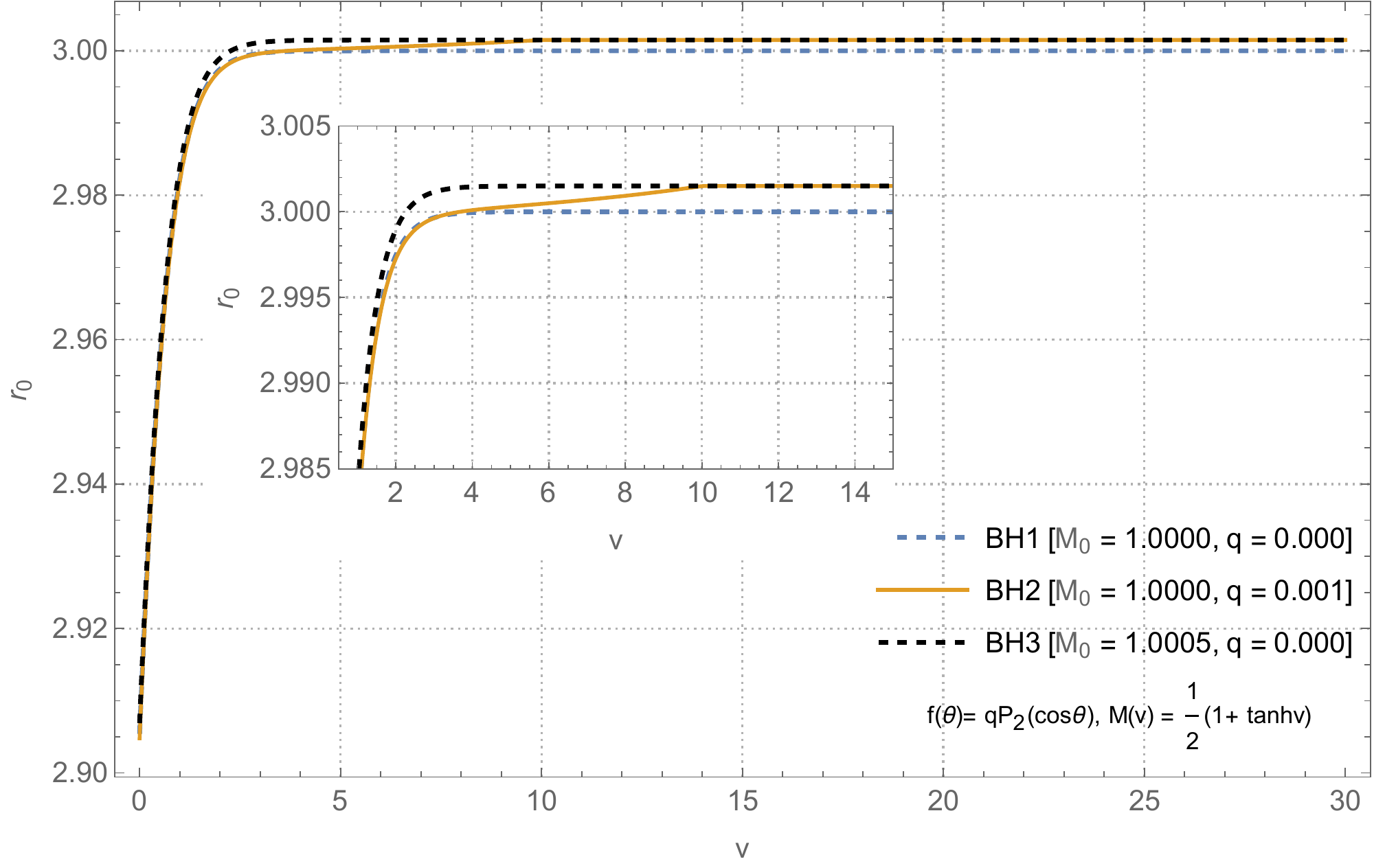}
			\caption{The evolution of three distinct BH configurations from an instant $v<v_0$.}
			\label{fig:2b}
		\end{subfigure}
		\caption{The evolution of the photon sphere of a STVBH from an instant of time $v<v_0$, and the null shockwave hits the BH at $v_0=10$, with $M(v)=\dfrac{1}{2}(1+ \tanh(v)), v_f=30$.The insets show close ups of the main plots.} 
		\label{STVBH2fig}
	\end{figure} 
	Putting $q=0$ or $v<v_0$ in \eqref{goryequation}, we recover the result for a bald Schwarzschild-Vaidya black hole reported in \cite{Mishra:2019trb}. To study the behaviour of the photon sphere of an STVBH, \eqref{goryequation} must be solved numerically subjected to two boundary conditions. It is physically reasonable to assume that at very late time, the dynamical black hole's  photon circular orbit will attain the value given by \eqref{bc}. We therefore take the following two future boundary conditions at a sufficiently late time $v_f$, namely, $r_0(v_f)= 3M_0+3q/2$ and $\dot{r}_0(v_f)=0$. We also choose to model the growth of the mass parameter using a smoothly growing function like
		\begin{align}
			&	M(v) = \dfrac{M_0}{2} \left[1 + \tanh(v)\right]. \label{mtanh} 
		\end{align}
	
	The evolution of the photon sphere can now be studied in two distinct ways: the STVBH is formed when a null shockwave implants some supertranslation hair on a bald Vaidya black hole at some time $v=v_0$. So we can either track the evolution of the photon sphere from an instant of time $v>v_0$ \textit{(Case I)} or from $v<v_0$ \textit{(Case II)}. We plot the results showing the evolution of the photon sphere of an STVBH for Case I in FIG \eqref{STVBH1FIG} and Case II in FIG \eqref{STVBH2fig} for the mass function given by \eqref{mtanh}.  
	
	\textit{Case I $(v>v_0)$:} In this situation, the black hole under investigation has already been implanted with supertranslation hair and its photon sphere is being studied as it accretes matter. Operationally, this amounts to dropping the term multiplied by the delta function, setting the step function to unity in \eqref{goryequation} and then solving it. In FIG \eqref{fig:1a}, we show the evolution of the photon sphere for various values of the supertranslation hair $q$. For $q=0$, we recover the result for a bald Vaidya black hole as expected. For small non-zero values of $q$, the photon sphere of a supertranslated BH evolves with a distinct trajectory but in a manner similar to its bald counterpart. This is demonstrated more concretely in FIG \eqref{fig:1b} where we study the photon sphere of three distinct BH configurations, labelled BH1, BH2 and BH3 respectively, side by side. BH1 is a bald BH whose mass saturates to $M_0=1$ and its photon sphere therefore naturally saturates to $r_0(v_f)=3.0$. However, BH2 is a supertranslated BH with $M_0=1,~ q=0.005$ and its photon sphere saturates to a value given by \eqref{bc}, viz, $r_0(v_f) = 3.0075$. Now if we consider the evolution of a black hole (BH3) with $M_0=1.0025,~ q=0.0$, we find that its photon sphere not only saturates to $r_0(v_f)=3.0075$ (as it should) but the evolution of its photon sphere also coincides with that of BH2, a supertranslated BH! Therefore, it seems that one might not be able to definitively say if one has observed a dynamic supertranslated BH or spotted a bald Vaidya BH with a different ADM mass just by ``looking" at the evolution of the photon orbit if the observation is made after the hair has been implanted.
	
	\textit{Case II $(v<v_0)$:} Physically, in this situation, we are looking at the black hole from an instant $v$ before the shockwave has hit the horizon at $v_0$. So, while solving \eqref{goryequation}, we keep the terms multiplied with the delta function that is triggered at $v=v_0$ along with the step function. We can therefore study the entire dynamical phase of the STVBH. While solving \eqref{goryequation}, we restrict ourselves to small values of $q$ and approximate the step function and the delta function with the error function, $S(v-v_0)=\mathrm{erf}(k(v-v_0))/2+1/2$, and its derivative respectively with $k=15$ for $v_f=30$.
	In FIG \eqref{fig:2a}, we plot the evolution of the photon sphere for various values of the supertranslation hair. To understand the scenario in a heuristic manner, we again compare the evolution of three distinct black hole configurations. Like before, BH1 and BH3 are bald black holes that evolve as explained earlier. Now, although BH2 starts out as a bald black hole with $M_0=1$, it gets endowed with the supertranslation hair due to the null shockwave. If there were no implantation of hair, the photon sphere of BH2 would have evolved like that of BH1. However, in practice, the manner in which the photon sphere of BH2 evolves changes drastically compared to BH1; and after a brief period of distinct evolution, it starts to coincide with that of BH3 as it saturates to a value given by \eqref{bc}. Of course, this is a rather qualitative picture but it provides an idea regarding how one may differentiate a supertranslated black hole from a bald black hole by the evolution of its photon sphere. It is worth noting that the null shockwave is a mathematical idealization. Therefore to accurately simulate the behaviour of a supertranslated BH, one needs to come up with physically realistic models which can implant supertranslation hair on a black hole.
	\paragraph{Discussion.}
	The behaviour of the photon sphere of dynamical black holes changes over time. We aimed to study the evolution of the photon sphere of a supertranslated Vaidya black hole (STVBH) to understand whether we can detect the presence of supertranslation hair through standard tests of general relativity that involve the bending of light due to strong gravitational fields. We find that the answer to this question is rather subtle.	The evolution of a photon sphere of a bald black hole that gets implanted with supertranslated hair in due course of time differs from that of an unperturbed bald black hole. This change in the photon sphere of a supertranslated black hole, however, can only be observed for a time period before the radiation (implanting the supertranslation hair) hits the horizon. Therefore it is reasonable to expect that this effect will give rise to a shadow that is distinguishable from a Vaidya black hole. Studying the evolution of the photon sphere is but a first step in this direction. Our analysis also indicates that if one studies the behaviour of the photon sphere of a black hole much after it gets implanted with the hair then one may not be able to distinguish it from a bald black hole. An analysis of the shadow of such a configuration will provide a definitive answer. But the computation of the shadow of an STVBH itself is quite difficult. Usually, if the angular and radial parts of the first integral of motion for null geodesics are separable by introducing the Carter constant then one can study the shadow using a semi-analytic approach \cite{Mishra:2019trb}. Unfortunately, the metric under consideration is not amenable to such treatment (since the $g_{vv}$ and $g_{v \theta}$ components of the metric depends on $\theta$ through the supertranslation field). Usually, the shadow of a static black hole with a metric where the null geodesic equations cannot be decoupled (for example, see \cite{Hu:2020usx}) is studied numerically using backward ray-tracing algorithms (\cite{Cunha:2016bjh} describes one such publicly available code). At present, algorithms for studying dynamical black holes are not available to the best of our knowledge. Hence it would be fruitful to develop a numerical method that can tackle spacetimes that do not possess energy as a conserved quantity, that is, dynamical spacetimes, and generate the evolution of its shadow. We humbly leave this issue for the future.

	\paragraph{Acknowledgments.}
	S. K. thanks Pavan Chakraborty and Anjan Ananda Sen for their valuable comments that led to the question addressed in this paper. We acknowledge very useful communications with  Geoffrey Comp{\`e}re and correspondence with M. M. Sheikh-Jabbari. S.S. also thanks Ayon Tarafdar, Indranil Chakraborty and Divyesh N. Solanki for useful discussions. Research of S.B. and S.S. is supported by DST-SERB, Government of India under the scheme Early Career Research Award (File no.: ECR/2017/002124) through the project titled \emph{``Near Horizon Structure of Black Holes"}.

	
	%

\end{document}